\documentclass[useAMS,usenatbib]{mn2e}
\usepackage{epsfig,natbib}
\usepackage{amssymb}
\usepackage{graphicx}
\usepackage{placeins}
\usepackage{times}
\usepackage{epsfig}
\usepackage{amsmath}

\def\dv90{\mbox{$\Delta v_{90}$}}
\def\ztau50{\mbox{$z_{\tau50}$}}
\def\sigmaem{\mbox{$\sigma_{\mathrm em}$}}
\def\halpha{\mbox{H$\alpha$}}
\def\hbeta{\mbox{H$\beta$}}

\def\lya{\mbox{Ly$\alpha$}}

\def\lesssim{\mathrel{\hbox{\rlap{\hbox{\lower4pt\hbox{$\sim$}}}\hbox{$<$}}}}
\def\gtrsim{\mathrel{\hbox{\rlap{\hbox{\lower4pt\hbox{$\sim$}}}\hbox{$>$}}}} 

\newcommand{\pkpc}{\ensuremath{\rm kpc^{-1}}}
\newcommand{\kms}{\ensuremath{{\rm km\,s^{-1}}}}
\newcommand{\zabs}{\ensuremath{z_{\rm abs}}}
\newcommand{\zem}{\ensuremath{z_{\rm em}}}
\newcommand{\vrel}{\ensuremath{v_{\rm rel}}}
\newcommand\ion[2]{#1{\sc #2}}
\newcommand\fion[2]{$[$#1{\sc #2}$]$}

\title[Metallicity has followed local gravitational potential of galaxies since z=3]
{Metallicity has followed local gravitational potential of galaxies
since $z=3$
}

\author[P. M{\o}ller and L. Christensen]{
P. M{\o}ller$^{1}$\thanks{E-mail: pmoller@eso.org}
and
L. Christensen$^2$\\
$^1$European Southern Observatory, Karl-Schwarzschildstrasse 2, 85748
Garching bei M\"unchen, Germany\\
$^2$Dark Cosmology Centre, Niels Bohr Institute, University of Copenhagen,
Juliane Maries Vej 30, 2100 Copenhagen O, Denmark
}

\begin{document}


\pagerange{1 -- 1} \pubyear{2011}

\maketitle


\begin{abstract}
The MZ relation between stellar mass (${\rm M}_*$) and metallicity (Z) of
nearby galaxies has been described as both a global and local property, i.e.
valid also on sub-galaxy scales. Here we show that Z has remained a local
property, following the gravitational potential, since $z=3$. In
absorption the MZ relation has been well studied, and was in place already
at $z=5.1$. A recent absorption study of GRB galaxies revealed a close match
to Damped \lya\ (DLA) galaxies, surprising due to their vastly different
impact parameters and leading the authors to suggest that local metallicity
follows the local gravitational potential. In this paper we formulate an
observational test of this hypothesis.
The test, in essence, forms a prediction that the velocity dispersion
of the absorbing gas in galaxy halos, normalized by the central velocity
dispersion, must follow a steep log scale slope of -0.015 dex \pkpc\ as a
function of impact parameter out to at least 20-30 kpc.
We then compile
an archival data and literature 
based sample of galaxies seen in both emission and absorption
suitable for the test, and find that current data confirm the hypothesis
out to 40-60 kpc.
In addition we show that the distribution of the velocity
offsets between \zem\ and \zabs\ favours a model where DLA systems are
composed of individual sub-clouds distributed along the entire path through
the halo, and disfavours a model where they are one single cloud with a
bulk motion and internal sub-structure.  

\end{abstract}


\section{Introduction}
The mass-metallicity (MZ) relation for galaxies is well documented and
has been observed both in emission \citep{tremonti04,maiolino08} and in
absorption \citep{christensen14,augustin2018,rhodin2018}. The emission MZ
relation, as originally determined from luminosity-selected galaxy samples,
describes a relation between the total stellar mass (${\rm M}_*$) of a
galaxy and its emission line metallicity (Z) determined from integrated line
fluxes across the central part of the galaxy. I.e. it was discovered as
a relation between global properties of the galaxies. Later studies,
with improved resolution, have shown that on sub-galaxy scales similar
relations govern the local metallicity in galactic discs, where it
correlates with the local stellar mass density
\citep{moran2012,sanchez2013}.  Those surveys also found that 
metallicity gradients are ubiquitous both in the nearby
universe \citep{sanchez2014,belfiore2017,sanchezM2018} and at high
redshifts \citep{wuyts2016}.

The MZ relation of absorption selected galaxies was initially identified
via a relation between the `velocity width' \dv90 \citep{prochaska97} of
the absorption lines and the absorption line metallicity ([M/H])
\citep{wolfe1998,ledoux06, neeleman13, som2015}. Both quantities are 
measured in a single pencil-beam along a sightline through the
circum-galactic medium (CGM) of a galaxy, its halo and 
in rare sightlines through the galaxy itself. The
sightline is defined by the position of an unrelated background quasar and
is therefore randomly chosen. Starting with \citet{ledoux06} it is now
customary to take \dv90\ to be a proxy for ${\rm M}_*$, and the underlying
assumption is here that \dv90\ and [M/H], both determined locally at a random
impact parameter $b$ (the projected distance between the pencil beam and
the galaxy centre), are close enough to the central values that the relation
is still identifiable. This assumption was vindicated by \citet{moller13}
who determined a procedure to calibrate the \dv90-[M/H] relation to an
${\rm M}_*$-[M/H] relation, and thereby to connect
it directly to the \citet{maiolino08} emission MZ relation. The
absorption relation includes an unknown term, $C_{[M/H]}$, describing
the average difference between emission metallicity and absorption
metallicity.  Subsequently \citet{christensen14} showed
that this term could be understood as a product $b\times \Gamma$, where
$\Gamma = -0.022\pm0.004$ dex kpc$^{-1}$ is the average metallicity
gradient of absorption selected galaxies.

\begin{figure}
\vskip -0.2 cm
\includegraphics[width=8.5 cm]{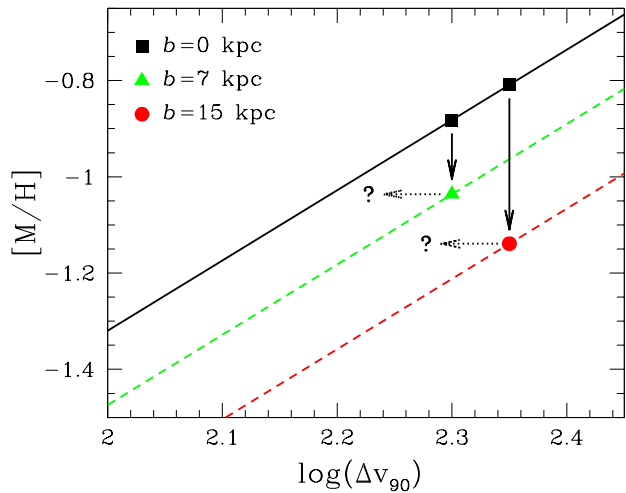}
\vskip -1.8 cm
\caption{Simple illustration of the \dv90-[M/H] relation and the effect
of different impact parameters, $b$. The full black line represents a
zoom on a small section of an
idealized \dv90-[M/H] relation with two galaxies observed at zero impact
parameter (black squares). The same galaxies observed at $b$ = 7 and 15
kpc are offset to lower metallicities because of the metallicity gradient.
If \dv90, as assumed, is tracing the local gravitational well depth, the
points would be expected also to move towards the left as illustrated by
the dotted arrows.}
\label{fig:IntroFig}
\end{figure}

For Damped \lya\
 (DLA\footnote{The terms DLA and sub-DLA refer to sightlines with 
 $\log N($H\,{\sc i}$) > 20.3$ and $19.0 < \log N($H\,{\sc i}$) < 20.3$ 
 respectively}) 
galaxies $b$ is found typically to be in the range 0-25 kpc
\citep{krogager17}, while for sub-DLAs $b$ is typically somewhat larger 
\citep{rahmani2016,rhodin2018}. Determination of $b$ requires that the
absorbing galaxy first must be found also in emission, a difficult task, and
$b$ is therefore known for only a tiny fraction of the known DLAs/sub-DLAs. 
The random and unknown value of $b$ for the majority of the absorbers,
combined with the metallicity gradient, result in a corresponding range of
unknown offsets towards lower metallicities in the \dv90-[M/H] relation.
For example a DLA at $b=10$ kpc will on
average have a measured metallicity $10 \times 0.022~{\mathrm dex}
= 0.22~{\mathrm dex}$ lower than the central metallicity, potentially
shifting the entire relation to lower metallicities and adding an extra
scatter reflecting the range of possible values of $b$. This
was thought to
be the source of some of the residual scatter present in the relation:
$\sigma_{\rm scatter}$([M/H]) = 0.38 \citep{moller13};
rms([M/H]) = 0.37 \citep{neeleman13}.

The effect of observing a galaxy at two different impact parameters is
graphically illustrated in Fig.~\ref{fig:IntroFig}. The green and
red dashed lines show how an idealized relation through the galaxy centres
(full black line) is offset towards lower metallicities for increasingly 
larger values of $b$. As mentioned above, \dv90\ has been used as a proxy 
for galaxy mass because it was thought it would trace the depth of the 
gravitational potential encountered.
However, at larger $b$ only the outer part of the potential well depth
will be probed by the sightline. Will this then also modify the relation?
A sample of DLAs with zero, or very small, impact parameters would be needed
to answer this question. Gamma Ray Bursts (GRBs) provide pencil-beam
line absorption data much like those of QSOs and their spectra usually
contain a DLA intrinsic to the host. GRB-DLAs have much smaller impact
parameters (in a sample study \citet{lyman2017} determined a median
of $b=1.0\pm0.2$ kpc) and as the GRB afterglow fades away, the host
galaxy light is not contaminated by an extremely luminous source.  
While the geometry for GRB host DLAs is slightly different (they only
have half the sightline compared to intervening DLAs), the smaller
impact parameter does make the comparison interesting.

\citet{arabsalmani15} analyzed a sample of GRB-DLAs with the goal to
compare the GRB-DLA \dv90-[M/H] relation to that of QSO-DLAs. Quite
surprisingly they reported that there was no offset between the
two relations,
i.e. the offset between the black line and the red dashed line in
Fig.~\ref{fig:IntroFig} was not seen. As explanation they suggested
that perhaps the real underlying relation is directly between [M/H] and
potential well depth, i.e. that the relation is local rather than global
(see their Fig. 2 and 4).
If that is the case then \dv90 will decrease with the exact rate to
compensate the drop in metallicity as $b$ increases. Graphically a decrease
in \dv90 is shown as the dotted arrows in Fig.~\ref{fig:IntroFig}. If the
relation is local then the dotted arrows
will continue left to the point where they again intersect the black
($b=0$) relation. In other words galaxies will move along
the relation, rather than across it, as $b$ changes, explaining why
the relation is so easily detectable despite the unknown impact
parameters. If the suggestion can be confirmed, then this will 
lend strong support to the reports of local MZ relations in resolved
studies of nearby (out to $z=0.05$) galaxy discs.  Even more
importantly it will extend those results out to high redshifts
and to sightlines far outside the visible parts of galaxies.

The result by \citet{arabsalmani15} cited above effectively suggests
a relation based on two
bins; one defined by the average impact parameter of GRB-DLAs, and another
defined by the average impact parameter of QSO-DLAs. The lack of knowledge
of the individual impact parameters prevented the authors of that paper
from confirming that the relation exists.
In this paper we aim to perform a direct test of the hypothesis
that the \dv90-[M/H] relation is the result of a more fundamental
local relation between the potential well depth and the
metallicity, both measured via neutral absorbing gas in the
CGM of a complete literature sample of DLAs and sub-DLAs.  
In \citet{christensen2019} we will elaborate further on the radial
dependence and compare to a set of dark matter (DM) halo profile
models.

\section{Sample properties}
\subsection{Sample definition}
\label{sect:sampledef}
For our test we require a sample of spectroscopically confirmed galaxy
counterparts to DLAs and sub-DLAs for which 4 parameters are known:
\sigmaem, $b$, \dv90\ and $N($H\,{\sc i}) representing the Gaussian
$\sigma$ of emission lines integrated over the host galaxy, impact
parameter, absorption line width and column density of neutral Hydrogen
in the absorption sight-line respectively. Implicitly this also means that
we know \zem\ and \zabs\ for each galaxy and each absorber. Early
emission line searches for DLA galaxies in emission were primarily aimed
at detection of \lya\ because the majority of known DLA systems had
redshifts $z>1.7$, where \lya\ is shifted into the optical wavelengths and
ground based optical observations therefore allowed extensive searches.
However, because of its resonant nature \lya\ was found to be
poorly suited for dynamical studies of the hosts. In this work we only
consider \halpha , \hbeta , \fion{O}{ii} and \fion{O}{iii} emission lines.
With current instruments those can now be found over a wide redshift range.  

\begin{table*}
\caption{QSO-DLA sample. Detailed referencing and account for the origin of
each entry in the table is provided in Appendix A where we also provide the
background quasar SDSS IDs. \zabs\ denotes the redshift of the absorber and
we use \ztau50\ when that is available. \zem\ is the redshift of the host
galaxy measured from Balmer or Oxygen emission lines.}
\begin{tabular}{lllcrccclcr}
\hline
ID & \zabs\ & \zem\ & $\log N($\ion{H}{i})&b&
$\log ({\rm M}_*/{\rm M_\odot})$ 
& \sigmaem\ & \dv90\ & \vrel\ & Metallicity & Refs.$^\ddagger$  \\
  &   &   &   & kpc &    &   \kms\   &  \kms\ &  \kms\ & [X/H]$^\dagger$ &
[X/H]  \\
\hline
0151+045    & 0.1602  & 0.1595  & 19.48 & 18.5 &$ 9.73\pm0.04$ & $50\pm20$
&  152 & $+180$     & $-0.04\pm0.15$ & (1) \\
2328+0022   & 0.65179 & 0.65194 & 20.32 & 11.9 &$10.62\pm0.35$ & $56\pm24$
&   92 & $-27\pm11$ & $-0.52\pm0.17$ & (2) \\
1323-0021   & 0.71612 & 0.7171  & 20.4  &  9.1 &$10.80^{+0.07}_{-0.14}$ & $101\pm14$ 
&  141 &$-171\pm12$ & $+0.40\pm0.3$  & (3) \\
1436-0051A  & 0.7377  & 0.73749 & 20.08 & 45.5 &$10.41\pm0.09$ & $99\pm25$
&   71 & $+36$ & $-0.05\pm0.12$& (4) \\
0153+0009   & 0.77219 & 0.77085 & 19.70 & 36.6 &$10.03^{+0.18}_{-0.08}$ &
$121\pm8$    &   58 &$+227$ & - & \\ 
0152-2001   & 0.77980 & 0.78025 & 19.10 & 54.  &   -   & $104\pm13$   &
33 & $-78$  & - & \\ 
1009-0026   & 0.8866  & 0.8864  & 19.48 & 39.  &$11.06\pm0.03$ & $174\pm5$
&   94 & $+32$ &$+0.25\pm0.06$ & (5) \\
1436-0051B  & 0.9281  & 0.92886 &$18.4$& 34.9 &$10.20^{+0.11}_{-0.08}$ &
$33\pm11$    &   62 &$-118$ & $-0.05\pm0.55$ & (6) \\
0021+0043   & 0.94181 & 0.94187 & 19.38 & 86.  &   -   & $123\pm11$   &
139 &  $-9\pm5$ & $+0.42\pm0.15$ & (7) \\
0302-223    & 1.00945 & 1.00946 & 20.36 & 25.  &$ 9.65\pm0.08$ & $59 \pm6$
&   61 &  $-1$ & $-0.54\pm0.13$ & (2) \\
2352-0028   & 1.03197 & 1.032   & 19.81 & 12.2 &$ 9.4\pm0.3$ & $125\pm6$
&  164 & $+40\pm5$ & $+0.17\pm0.13$ & (8) \\
\hline
2206-1958   & 1.919991& 1.9220  & 20.67 &  8.4 &$ 9.45\pm0.30$ & $93\pm21$
&  136 &$-210$&$-0.60\pm0.05$& (9) \\
1228-1139   & 2.19289 & 2.1912  & 20.60 & 30.  &   -   & $93\pm31$    &
163 &$+159$ &$-0.22\pm0.10$& (2) \\
1135-0010   & 2.2066  & 2.207   & 22.10 & 0.8  &   -   & $53\pm3$     &
168 &$-20\pm10$ &$-1.06\pm0.10$& (10) \\
2243-6031   & 2.3298  & 2.3283  & 20.67 & 26.  &$10.1\pm0.1$ & $158\pm5$
& 173 &$+135\pm12$ &$-0.91\pm0.05$& (9) \\
2222-0946$^x$   & 2.3542  & 2.3537  & 20.65 & 6.3  &$ 9.62\pm0.12$ &
$49.0\pm1.6$ &  174 &$+42\pm30$ &$-0.53\pm0.07$& (11) \\
0918+1636-1$^x$ & 2.412   & 2.4128  & 21.26 & $<2.$&   -   & $21\pm5$   &  344
&$-38\pm25$ &$-0.66\pm0.24$& (12) \\
1439+1117   & 2.41802 & 2.4189  & 20.10 & 39.  &$10.74\pm0.17$ & $303\pm12$
&  338 &$-77$ & $+0.20\pm0.11$ & (13) \\
0918+1636-2 & 2.5832  & 2.58277 & 20.96 & 16.2 &$10.33\pm0.08$ & $107\pm10$
& 288 &$+36\pm20$ &$-0.19\pm0.05$& (14) \\
2358+0149   & 2.97919 & 2.9784  & 21.69 & 11.8 &   -   & $47\pm9   $  &
135 &$+60$ & $-1.90\pm0.18$ & (15) \\
2233+1318   & 3.14930 & 3.15137 & 20.00 & 19.5 &$ 9.85\pm0.14$ & $23\pm10
$  &  228 &$-150$ &$-0.97\pm0.13$ & (16) \\
\hline
\hline
\end{tabular}
\flushleft
$^\dagger$~ For each absorber we have recomputed [M/H] to the solar
relative abundance as defined in \citet{decia2016} based on 
\citet{asplund2009}.  \\
$^\ddagger$~ Element used and column density references:
(1) [S/H]  \citep{som2015} the value given includes their ionization 
correction of -0.26 dex; 
(2) [Zn/H] \citep{berg2015}; 
(3) [Zn/H] \citep{moller2018}; 
(4) [Zn/H] \citep{meiring2008};
(5) [Zn/H] \citep{meiring2007}, they find an ionization correction of 0.15
dex but they do not apply it, nor do they add to the error but keep it as
0.06; 
(6) [Zn/H] \citep{meiring2008,straka2016} details given in Appendix
\ref{app:a};
(7) [Si/H] \citep{peroux2016};
(8) [Si/H] \citep{meiring2009a};
(9) [Zn/H] \citep{decia2016};
(10) [Zn/H] \citep{kulkarni2012}; 
(11) [Zn/H] \citep{fynbo2010}; 
(12) [Zn/H] \citep{fynbo13}; 
(13) [Zn/H] \citep{noterdaeme2008}; 
(14) [Zn/H] \citep{fynbo2011}; 
(15) [Zn/H] \citep{srianand2016}; 
(16) [Si/H] this work Appendix \ref{app:a}.\\
$^x$~ Targeted by the X-shooter survey, see
Sect.~\ref{sect:selectionfunction}.\\

\label{tab:sample}
\end{table*}

We carefully searched the literature for all such systems and have compiled
a complete literature sample of 21 absorption selected galaxies with the
required information (Table~\ref{tab:sample}). For several of the objects in
Table~\ref{tab:sample} not all the required parameters had been extracted
from the data in the original papers. For those we reprocessed the
original (or supplementary) data as detailed in Appendix \ref{app:a},
where we also
provide references to the original publications. One additional parameter
(the stellar mass, ${\rm M}_*$) is also relevant for the present purposes.
When available we have also recorded this in Table~\ref{tab:sample}.
The sample has been collected from programmes utilizing a wide range
of instrumentation, observing strategies and reduction techniques
each adapted to fit the wide range of purposes and goals of the original
studies. The original publication formats of those data are therefore
very inhomogeneous, and it has been necessary to re-process several
observables into a homogeneous data-set.

Our sample is well spread over the observable parameter space and in
order to assess dependencies on the various parameters we shall therefore
also consider sub-samples as defined here.
In total our sample contains 11 hosts at \zem\ below 1.05 (the low redshift
sample) and 10 hosts at \zem\ above 1.90 (the high redshift sample).
It contains 11 DLAs (the DLA sample), 9 sub-DLAs and 1 absorber with
even slightly lower $\log N$(H\,{\sc i}). We shall refer to the 10
low $\log N$(H\,{\sc i}) hosts together as the ``sub-DLA'' sample.  
Unfortunately there is a strong tendency for DLAs and sub-DLAs in our
sample to cluster in the `high redshift' and `low redshift' sample
respectively.  This causes some degeneracy regarding dependency on
$z$ and $\log N$(H\,{\sc i}).

For 15 of the hosts there are stellar masses known from SED fitting.
For 7 of those $\log ({\rm M}_*/{\rm M_\odot})$ is in the range
$9.4-10.03$ (the low mass sample), the other 8 have
$\log ({\rm M}_*/{\rm M_\odot})$ in the range $10.1-11.06$ (the
high mass sample).

Historically the offset between \zabs\ and \zem\ was used as the first
tracer of the dynamical relation between the host galaxy and its CGM
\citep[e.g.][]{warren1996, christensen2005},
but its use was complicated by the lack of a unique definition of
\zabs . Most DLA absorbers have complex multi-component structures
spanning in excess of 100 \kms\ or even several times that, and some
convenient value within this range has often simply been chosen. Until
now there has been no strong motivation to formulate a unique
definition. With a rapidly increasing sample size of DLA galaxies with
well defined emission redshifts this situation has changed. A useful
definition should represent a well defined ``average'' of the complex and
since the other absorption related dynamical tracer (\dv90) is formed by
cuts in the distribution of optical depth of the low-ion phase, we
shall here follow \citet{moller2018} and use (where possible) the
redshift of median optical depth (\ztau50), i.e. the redshift where
there is 50\% of the low-ion optical depth on either side. We then
define the relative velocity offset (\vrel ) as ($\zabs-\zem$) converted
to velocity, i.e. it is positive when the absorption occurs at higher
redshift than the emission.

To obtain a uniform set of impact parameters, $b$, we extract those in
arcsec from the original papers and convert them to kpc assuming a
flat cosmology with $H_0=70.4$ \kms\ Mpc$^{-1}$,
$\Omega_{\Lambda}=0.727$ \citep{komatsu2011}.  
In this cosmology, a 1-arcsec transverse separation on the sky 
corresponds to 7.6 kpc at $z=0.8$ and 8.3 kpc at at $z=2.4$.

For reference we also, in column (10) of Table~\ref{tab:sample}, list 
absorption metallicities when available. In order to obtain a 
homogeneous sample we have searched the literature for the measured 
total metal column densities and have then 
computed metallicitites based on solar abundances given in 
\citet{asplund2009}. In particular we also use the choices
used by \citet{decia2016} (their Table 1). We only consider
metallicities based on Zn, S and Si as provided in the references listed
in column (11) of Table~\ref{tab:sample}. Metallicities of our sample 
span the range from $-1.90\pm0.18$ to $+0.42\pm0.15$ i.e. 2.3 dex.

\subsection{The DLA galaxy selection function}
\label{sect:selectionfunction}

Any sample of high redshift galaxies contains intrinsic biases which are
a function of the exact way the galaxies are identified. The selection
function of DLA galaxies contains three components. First there is the
selection of the DLA absorption line which is a selection via \ion{H}{i} 
absorption
cross-section. The probability that a given galaxy is selected as absorber
scales directly with the area that the DLA gas of that galaxy covers on the
sky and that area scales, via the Holmberg relation, with the luminosity
of the galaxy 
resulting in a flat selection function over a wide span of luminosities
\citep[see figure 10 of][]{krogager17}. This is
in sharp contrast to luminosity selected galaxy samples which, by
construction, have a luminosity cutoff. Other than the
cross-section weighting, the selection of the individual DLAs forms
a random sampling onto any scaling relation (e.g. MZ and \dv90-[M/H]
relations).

The second component of the selection function is the method by which one
seeks to detect emission from the host. The first successfully used methods
were the \lya\ narrow band imaging technique \citep{moller1993}
and the Lyman Break broad band imaging selection \citep{steidel1995}.
HST imaging with follow-up spectroscopy of candidates
\citep[e.g.][]{lebrun1997,warren2001} and later both optical and IR
IFUs (e.g. PMAS \citep{christensen2007} and SINFONI \citep{peroux2011a}) 
were used. Latest, also ALMA has proven to be a powerful tool to identify
hosts via molecular emission lines \citep{neeleman2016,klitsch2018}. In
an interim period, before the advent of the new generation of powerful
data-cube instruments, a single slit triangulation method was used with
significant success \citep{moller2004,fynbo2010,krogager17}. Each method
has its limitations, notably in terms of the field covered.

The last component of the selection is the target selection. Initially the
selection of targets was either random or designed to cover the known
parameter space evenly \citep[see e.g. figure 1 of][]{warren2001}. Later
it was shown that luminosity scales with metallicity of the DLA,
and that it therefore is more telescope time efficient to 
only select DLAs above a given metallicity threshold since they are likely
to have brighter hosts. Such a biasing towards higher metallicity in some
of the later samples is equivalent to a normal luminosity bias, i.e. it
skews the
underlying flat selection towards mostly brighter galaxies. As above, this
does not affect the random sampling of the underlying scaling relations, it
simply means that just like for flux limited galaxy samples, only the
brighter end of the relations are studied. The shape of the relations
will not be biased. 
However, the target does need to be inside
the field of view (FOV) of the observation in order to be detected.

For all surveys based on imaging and IFU cubes with sufficiently large
field, the target will always be inside the FOV. Surveys which only partly
cover the possible impact parameter range, e.g. slit
triangulation, will inherently contain an additional bias towards more
easily finding hosts at smaller impact parameter and maybe missing hosts
at larger impact parameter. Scaling relations containing impact parameter
could be affected by such a detection bias. In our sample such a bias is
only present for the two hosts identified via the X-shooter triangulation
survey (0918+1636-1 and 2222-0946). A third host (0918+1636-2) was not in
the targeted sample of that survey but was identified serendipitously
during the search for the lower redshift host. The two sample members are
marked in Table~\ref{tab:sample}.
The strength of the effect was computed by \citet{fynbo2010} who find
that ``slightly above 90 per cent of the galaxy centres are covered by at
least one slit''. Therefore, in the complete triangulation survey
(10 targets), less than 1 host is expected to be missed due to large
impact parameter.  In our sample defined here the effect is $0.2$ target
and therefore negligible.

\section{Results}

\subsection{Mass-metallicity relation, local or global?}
\label{sect:slopetest}
 
The first question we examine is the hypothesis that the metallicity
measured along a pencil beam is related to how deep the sightline dips into
the gravitational potential, rather than to the total stellar mass of the
galaxy \citep[as illustrated 
by the sightline trace in Figure 4 of][]{arabsalmani15}.
In order for this to explain why GRB-DLAs and QSO-DLAs follow exactly the
same relation, despite their vastly different distribution on impact 
parameters, there must be a gradient of \dv90\ outwards through the galaxy 
halo which exactly cancels the metallicity gradient (as illustrated in 
Fig.~\ref{fig:IntroFig}). The metallicity gradient for DLA galaxies has been
reported as $\Gamma = -0.022\pm0.004$ dex kpc$^{-1}$ \citep{christensen14}
and $\Gamma = -0.022\pm0.001$ dex kpc$^{-1}$ \citep{rhodin2018}, and the 
gradient of [M/H] vs log(\dv90) is 1.46 \citep{ledoux06,moller13}. In
order to cancel each other out the gradient of log(\dv90) should be
$-0.022/1.46=-0.015$ dex kpc$^{-1}$.
The above forms a strong prediction which is based on a sample of 110
QSO-DLAs \citep{moller13} and a sample of 16 GRB-DLAs
\citep{arabsalmani15}. If our present, independent, sample is found to
follow this prediction, then the result from those three samples taken
together makes a strong case in favour of the tested hypothesis.

\begin{figure}
\vskip -0.0 cm
\includegraphics[width=8.5 cm]{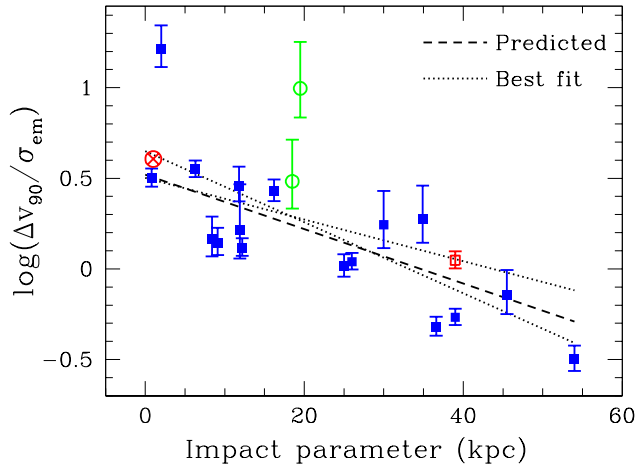}
\vskip -2.4 cm
\caption{Local dynamical state of the halo gas (\dv90) normalized to the
central value of \sigmaem, as a function of $b$ (blue squares). The green
open circles are objects where the emission line region is offset by a few
kpc from the galaxy centre and where \sigmaem\ consequently could
underestimate the central velocity dispersion. The red open square is AGN
dominated. The black dashed line is the hypothesis being tested while the
two dotted lines are two fits to the data (see text). The red circle with
an X is the average point for GRB-DLAs and was not included in the fits.
}
\label{fig:slopetest}
\end{figure}

\begin{figure}
\includegraphics[width=8.5 cm]{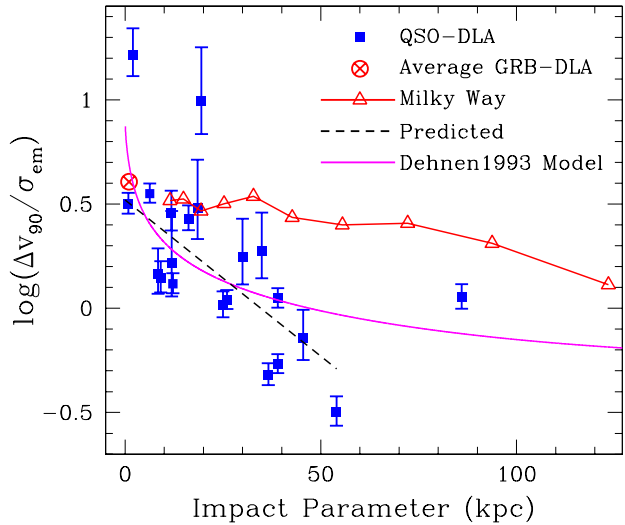}
\vskip -1.5 cm
\caption{Average projected velocity dispersion profile of our absorption
selected sample (solid blue squares) compared to the projected velocity
dispersion profile of our own galaxy (open red triangles). Predicted
slope and average GRB-DLA are included as in Fig.~\ref{fig:IntroFig}. We
also include an example model curve from \citet{dehnen1993}. DLA potentials
are seen to be better represented by steeper potentials than the MW.
}
\label{fig:MWtest}
\end{figure}

If we take \sigmaem\ to be a measure of the central velocity dispersion
of the gas, then we can use this to normalize \dv90\ for each pencil 
beam absorber\footnote{Normalizing the DLA sample to the ratio between 
\dv90\ and \sigmaem\ means that it is not normalized to unity, but
to the ratio between the two definitions. That ratio is, in the ideal 
case of a Gaussian profile, simply \dv90\ of a Gaussian line profile in
units of $\sigma$, which is $3.29$ or $\log(3.29)=0.517$.
(\dv90 / \sigmaem) is then a dimensionless observable which describes 
the dynamical state of the gas at $b$ relative to the centre.}.  
In Fig.~\ref{fig:slopetest} we therefore plot log(\dv90 / \sigmaem) vs. $b$
(blue squares), as well as the line with a slope of $-0.015$ and
intercept = $0.517$ (black dashed line) which is the prediction
we test. The two dotted lines are the minimum $\chi^2$ fit linear
relations. For a better visual impression we have limited the plot
to the inner 60 kpc but there is an additional single object at $b=86$ kpc 
(see Fig.~\ref{fig:MWtest}). The two dotted lines in
Fig.~\ref{fig:slopetest} are fits including (slope of $-0.011$) and
excluding
(slope of $-0.020$) this extra-large impact parameter object. The two green
points are objects for which we know that the emission line region is offset
from the centre of the galaxy by 3.8 and 2.1 kpc (for 0151+045 and 2233+1318
respectively, details provided in Appendix \ref{app:a}) and that
they therefore could be representing gas in smaller star-forming regions
rather than the entire galaxy. I.e. they might provide an under estimate
of the globally integrated \sigmaem, and as such could then become upwards
outliers.  In the figure one green point is seen to 
follow the general distribution while the other lies far above it.
The \sigmaem\ of the outlier, 2233+1318, is also found to be
unusually low compared to the stellar-mass Tully-Fisher relation
\citep{christensen2017},
but we shall here conservatively keep both in the sample.
The open red square is an AGN dominated DLA host (1439+1117, see Appendix
\ref{app:a}),
and in the paper reporting the discovery of this host \citet{rudie2017}
argue that the absorbing gas in this case is the result of an AGN driven
outflow. It
is of great interest to test if, in such cases, the kinematic properties
of the absorber are dominated by the outflow mechanism - or if they are
tracing the gravitational potential regardless of the outflow origin. It is
seen that this DLA lies precisely on the relation of the potential
well hypothesis.

We can obtain an additional point at small impact parameter if we also
consider GRB-DLAs. A complete literature sample of 10 GRB-DLAs with the
required dynamical parameters \dv90\ and \sigmaem\ was presented in
\citet{arabsalmani2018}. We do not know the exact impact parameters
for each, but GRB impact parameters are always small. Here
we simply compute the average $\dv90 /\sigmaem$, and use the median impact
parameter ($1.0\pm0.2$ kpc) 
found by \citet{lyman2017}.
One of the objects of that sample (GRB 090323A) has an extremely
large \dv90\ of 843 \kms, more than twice that of any object in our
QSO-DLA sample. In a separate study \citet{savaglio2012} conclude that
this system is caused by two separate galaxies and for this reason we
have removed this object before computing the average. The average
$\log(\dv90 / \sigmaem)$ of the remaining 9 is $0.61\pm0.06$ which is
included in Fig.~\ref{fig:slopetest} as a red circle with an X.
It was not included in the computation of the fits shown as dotted
lines. Including all 10 GRB-DLA in the average gives instead $0.70\pm0.08$
which would lie only insignificantly higher in the plot.  

It is immediately clear from Fig.~\ref{fig:slopetest} that the points,
out to an impact parameter of 60 kpc, follow the 
predicted
slope well. From this we conclude:
\begin{enumerate}

\item{there is an outward negative gradient of log(\dv90 / \sigmaem)}

\item{the slope of that gradient is in excellent agreement with the
hypothesis that this is the cause of
the reported
lack of difference between
the \dv90-[M/H] relations of QSO-DLAs and GRB-DLAs}

\item{the implication is that the MZ relation exists locally, i.e. that
both metallicity and \dv90\ follow the local gravitational potential. In
particular we find that this is true all the way back to $z=3.1$, and out
to larger distances than enclosed by galaxy disks.}

\end{enumerate}

In this section our aim was to test if our sample confirms the prediction
of the previous, independent samples. In Sect.~\ref{sect:Quantifying} we
shall return with a full mathematical outline of our current understanding
of the \dv90-[M/H] relation and in particular determine the observed
\dv90 gradient, $\Gamma_{\Delta v90}$.

\subsection{Absorption pencil beams: dynamical tracers of halo potentials}

\begin{figure}
\includegraphics[width=8.5 cm]{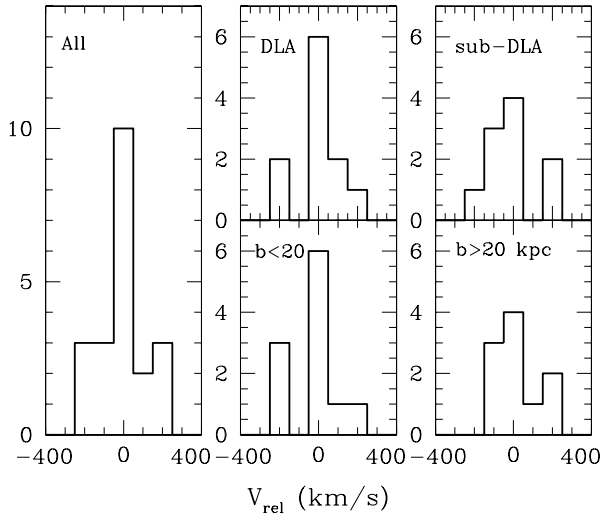}
\vskip -1.7 cm
\caption{Histograms of the distribution of \vrel. Left panel shows that
of the entire sample, middle and right panels present two different cuts
on the sample.
}
\label{fig:histogram}
\end{figure} 

In the previous section we have shown that the normalized \dv90\ of a sample
of DLA galaxies provide insight into the gravitational well profiles of
their halos, and by inversion, into their matter density profiles. Before we
exploit this further, the sample as reproduced in Fig.~\ref{fig:slopetest}
deserves a few words of clarification.  
The figure is a representation of the projected velocity dispersion profile
of the average potential well of the galaxies in our sample. Each point
represents the relative drop in projected velocity dispersion from the
centre to the given impact parameter, but they are all measured in
separate halos.  If all those galaxies were embedded in dark matter
halos of vastly different profile steepness, then those differences
would be expected to result in a large scatter of the individual points.
For the fits shown in Fig.~\ref{fig:slopetest} we used a method which
simultaneously determines the minimum $\chi^2$ fit and the internal scatter
of the data \citep{moller13}. The resulting scatter around the simple
linear fit is $0.26$ and $0.31$ excluding and including the absorber at
$b=86$ kpc respectively. This scatter is small compared to the full span of
the data points which is 1.7 dex, indicating that absorption selected
galaxies have DM halos with similar circum-galactic profile slopes.

In Fig.~\ref{fig:MWtest}, for comparison, we plot the projected velocity
dispersion for an example potential computed from the generalized model
by \citet{dehnen1993}. This model has two scaling parameters ($a$, the
scaling radius, and $\gamma$ describing the inner slope of the central
density profile) and it includes both the \citet{jaffe1983} and the
\citet{hernquist1990} models as special cases.  A rigorous and comprehensive
discussion of our DLA sample in the framework of those model potentials
is presented
in \citet{christensen2019}. Here we only point out that
the simplifying linear approximation we applied in the previous section to
investigate and resolve the relation between metallicity and potential is
well justified out to an impact parameter of 40-50 kpc. In that inner CGM
range the difference between the linear prediction and the \citet{dehnen1993}
model is well within the scatter (0.26 dex). The single sub-DLA point at
large impact parameter is clearly in disagreement with an extrapolation of
the approximated inner CGM linear slope, but it is in good agreement with the
\citet{dehnen1993} model at larger distances. A sample of securely
verified absorber/host pairs at larger
impact parameters is required to address the question of the outer halo
potential. For now we conclude that DLA galaxies in general have similar
density profile slopes in the inner CGM,
and that they likely are flattening at larger radii as predicted
by models.

Also in Fig.~\ref{fig:MWtest} we plot the projected velocity dispersion
profile of the Milky Way Galaxy (MW) from \citet{battaglia2005} (red
triangles) together with the DLA/sub-DLA data. To ease the comparison
we have normalized the MW data to its central value the same way we
have normalized our absorber
sample. In the representation here there is a significant difference
between the MW halo and those of most of the DLA galaxies.
This difference is related to differences in scaling radii and is
addressed in detail by \citet{christensen2019}.

\subsection{Sub-structure within, and distribution of, DLA absorbing clouds}

\begin{figure}
\includegraphics[width=8.5 cm]{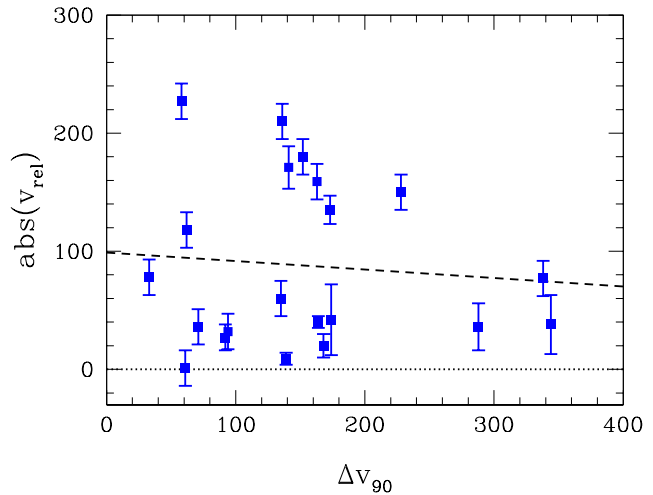}
\vskip -2.0 cm
\caption{Here we test if abs(\vrel) correlates with \dv90. A min-$\chi^2$
fit (dashed black line) finds a weak anti-correlation, but the data are
fully compatible with a slope of zero and no correlation.
}
\label{fig:vreldv90test}
\end{figure}

In Table~\ref{tab:sample} we also record \vrel, the emission vs. absorption
velocity difference, which we defined as positive in case the absorption
has the higher redshift. Let us consider two simple, opposing views
of the absorbing clouds. First, one could imagine that an absorber is a
single unity which has a bulk motion velocity, and within which the
individual sub-components reflect random motions relative to this bulk
motion. In this case \vrel\ represents the bulk motion and we would expect
it to roughly trace the halo potential, i.e. on average it should be
larger at small impact parameters and smaller further out. The velocity
width \dv90\ would then be an intrinsic property of the complex and would
not be strongly coupled to the halo potential.  The other simplified view
is that all the individual sub-components of a DLA system
represent independent absorbing clouds spread along the pencil beam through
the halo. In this case the combined velocity width of the complex (\dv90)
should be strongly coupled to the potential, and \vrel\ would simply
represent the average of the ensemble, i.e. it would be stochastic of
nature and should not be strongly linked to the potential.

All values of \vrel\ in our sample fall in the range $-210$ to $+227$ \kms\
and in Fig.~\ref{fig:histogram}, left panel, we show the histogram of their
distribution. Half of the velocities are within a narrow range of $\pm50$
\kms, the other half has a flat distribution over the full range $\pm220$
\kms. The distribution is symmetric around zero. In the middle and right
panels we use cuts on the sample to test if any obvious evidence is at
hand for a coupling to the halo potential. DLAs are known to typically be
closer to their host galaxies
while sub-DLAs typically are found at larger distances, so indirectly this
cut traces inner vs outer halo. We also make a cut directly on the impact
parameter at $b=20$ kpc. In case \vrel\ would follow the potential we would
expect the central peak of the distribution to be higher, and the
distribution to be narrower, in the right
panel than in the middle panel. The sample size is very small and at
present all of the DLA, sub-DLA, inner and outer samples are
statistically consistent with being identical. There is certainly no
evidence that the peak around $\pm50$ \kms\ is stronger in the outer parts
(rightmost panels).

In the previous sections we have shown that \dv90\ is strongly linked to
the potential, so we can make an additional test and ask the question if
\vrel\ is positively correlated with \dv90. For this we consider the
absolute value, abs(\vrel), and in Fig.~\ref{fig:vreldv90test} we plot this
against \dv90. We only have errors for 9 of the 21 values in
Table~\ref{tab:sample}, but redshifts
are in general easy to determine, so we do not expect any of the errors
to be large. Of the 9 values we determine the median error to be
12 \kms, so for the purpose of Fig.~\ref{fig:vreldv90test} we assign
an error of $\pm15$ \kms\ to the remaining 12. A minimum $\chi^2$ fit results
in a weak negative slope of $-0.072$ (dashed line in
Fig.~\ref{fig:vreldv90test}) for an intrinsic scatter of 67 \kms. The
data are also fully consistent with a slope of $0.0$, i.e. with no
correlation.

We conclude that \vrel\ of our sample is symmetrically
distributed between extremes of $\pm220$ \kms\ but 50\% of the sample is
found inside a narrow range of $\pm50$ \kms. Contrary to \dv90, \vrel\
does not correlate strongly with the halo potential. In terms of our simple,
idealized DLA model, those results favour that the multi-component structure
of DLAs represent ensembles of independent absorbers spread widely along
the pencil beam through the halo. In our data we cannot directly measure
the path length along which the absorbers are distributed, but in figures
3 and 6 of \citet{bird2015} it is seen that simulations of DLAs show how 
sub-components are distributed through the halo and that the physical path
length scales directly with \dv90 in agreement with our conclusion. This also
means that other properties of the DLAs, such as e.g. average metallicity
and distribution of metallicity on individual sub-components, represent the
averages and distributions taken through the entire halo.

\subsection{Dependence on redshift and/or \ion{H}{i} column density}

\begin{figure}
\includegraphics[width=8.5 cm]{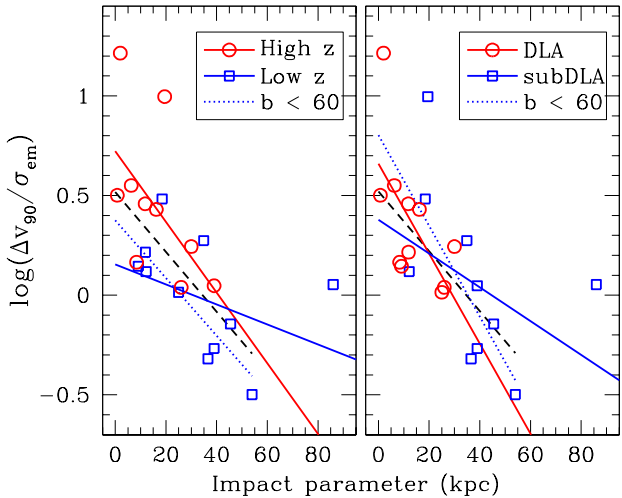}
\vskip -1.7 cm
\caption{Comparison of sub-samples as defined in Sect.~\ref{sect:sampledef}.
{\it Left:} Low redshift (blue squares and full line) vs. high redshift
(red circles and full line). {\it Right:} Low column density (blue) vs.
high column density (red). In both panels it is seen that the red points
and fits follow a steeper potential than the blue. However, excluding the
single point at $b=86$ kpc (blue dotted lines) the red and blue slopes are
seen to agree well. }
\label{fig:redshift}
\end{figure} 

As described in Sect.~\ref{sect:sampledef} there are some `natural cuts' on
our sample which can be used to obtain a first gauge of possible parameter
dependencies. First, for observational reasons (space vs. ground based
access to \lya) there is a natural division into a low and a high redshift
sample: \zem\ below 1.05 (11 hosts) and \zem\ above 1.90 (10 hosts).
The two sub-samples are shown in Fig.~\ref{fig:redshift}, left panel.
The dashed black line is again the same as in previous figures, while the
two full lines represent the best fit to the individual sub-samples.
It is seen that the high redshift sample favours a steeper potential while
the low redshift sample favours a flatter. However, recalling that the
model DM halo profile shown in Fig.~\ref{fig:MWtest} shows strong
flattening at impact parameters larger than 40-60 kpc, we repeat the fit
now excluding the single point at $b=86$ kpc, and show the fit as the dotted
blue line. The slopes of the full red and dotted blue lines are now
seen to be in very good agreement. In the right panel we plot the DLA
sample (red) and the sub-DLA sample (blue). The conclusion here is identical
to that of the left panel; when we exclude the $b=86$ kpc point we find an
excellent agreement between the low and high column density points.

In conclusion there is no evidence for neither
redshift dependence, nor \ion{H}{i} column density dependence, of the
steepness of the inner CGM slope (here taken to mean the average slope out
to a radius of 50 kpc) in DLA galaxy halos. On the contrary the evidence
is that they are very similar across those two parameters.

\begin{figure}
\includegraphics[width=8.5 cm]{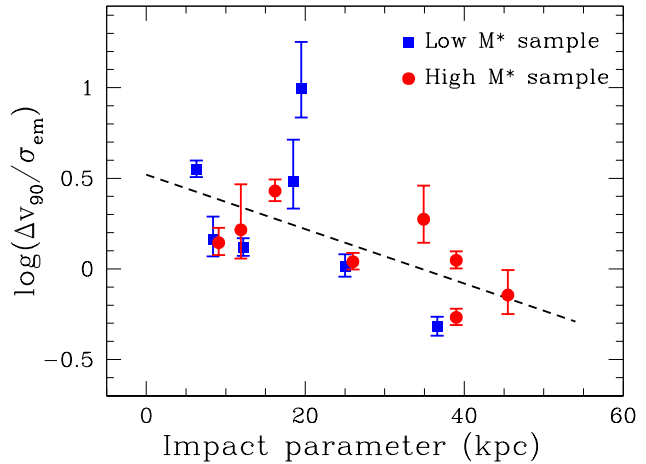}
\vskip -2.4 cm
\caption{As Fig.~\ref{fig:slopetest} but here only the 15 galaxies for which
the stellar mass is known. Both the low (blue squares) and the high (red
dots) stellar mass sub-samples are seen to lie on the same relation.} 
\label{fig:stellarmass}
\end{figure} 

\subsection{Dependence on stellar mass}

In Figures \ref{fig:slopetest} and \ref{fig:MWtest} we used \sigmaem\ to
normalize the \dv90\ measurements. Since \sigmaem\ is known to scale
with ${\rm M}_*$ \citep{christensen2017} this normalization takes out
the stellar mass to first order. However, in case galaxies of different
mass would sit in potentials with very different slopes, this could
still be visible as a secondary effect. For this reason we show, in
Fig.~\ref{fig:stellarmass}, a figure similar to Fig.~\ref{fig:slopetest}
only here we plot the low mass sub-sample as blue squares and the high
mass sub-sample as red dots (all sub-samples are defined in
Sect.~\ref{sect:sampledef}).
Low mass galaxies are in general smaller than galaxies with higher mass,
so it is unsurprising that we find a trend that lower mass galaxies
in general are found to have smaller impact parameters than the higher mass
galaxies. There may be a weak trend suggested that the low mass galaxies
have steeper slopes than high mass galaxies but a larger sample is required
in order to make a statistically valid statement about this. At present
the two sub-samples are consistent with following the same relation to
within the scatter.

\section{Discussion}
\subsection{The detailed \dv90-[M/H] relation}
\label{sect:The_detailed}

We are now able to piece together all of the various dependencies that
go into the seemingly simple relation between metallicity and \dv90, a
relation which was first reported by \citet{wolfe1998} who interpreted
it as due to rapidly rotating disks with metallicity gradients.
\citet{ledoux06} enlarged the sample size to 70 and suggested instead
that the relation was an absorption based version of the MZ relation,
a suggestion confirmed by
\citet{christensen14} who identified both the underlying MZ relation,
but who also observationally confirmed the existence of a metallicity
gradient. \citet{ledoux06} further found evidence for an evolution
with redshift, and in this paper we have now shown that also the
\dv90\ parameter displays a negative radial gradient. Here we will
formulate all this into a single description.

We start by considering a sightline through the centre of a galaxy.
Such a sightline will have $b=0$ and it will measure the central
values of the metallicity and \dv90\ which we will name
[M/H]$_{\rm c}$ and $\Delta v_{90,c}$ respectively. We can then write
the \dv90-[M/H] relation for sightlines through the centres of
galaxies as
\begin{equation}
\label{eq1}
{\rm [M/H]_c} = \alpha_0 \log( \Delta v_{\rm 90,c}) + \beta
\end{equation}
Equation~\ref{eq1} is valid at $z=0$, but for a given \dv90\ the
metallicity of the galaxy was lower in the past. \citet{moller13} and
\citet{neeleman13} provide expressions for that redshift evolution
but here we shall simply represent the evolution by a generic
function $ze(z)$, and
\begin{equation}
\label{eq2}
{\rm [M/H]_{c,z}} = \alpha_0 \log(\Delta v_{\rm 90,c}) + \beta - ze(z)
\end{equation}
is then valid for sightlines through centres of galaxies at any 
redshift. In order to generalize this further to a sighline at non-zero
impact parameter $b$ we follow \citet{christensen14} and write
\begin{equation}
\label{eq3}
{\rm [M/H] = [M/H]_c + \Gamma_{\rm [M/H]}} ~ b
\end{equation}
for the dependence of metallicity on the radial distance, and similarly
we write
\begin{equation}
\label{eq4}
\log({\rm \dv90}) = \log(\Delta v_{\rm 90,c}) + \Gamma_{\Delta v90} ~ b
\end{equation}
for the dependence of velocity dispersion on the radial distance.
Combining equations \ref{eq1} through \ref{eq4}, and assuming that the 
two gradients $\Gamma_{\rm [M/H]}$ and $\Gamma_{\Delta v90}$ do not 
depend on redshift, we can now write the full relation as
\begin{equation}
\label{eq5}
{\rm [M/H] = \alpha_0} (\log(\dv90) - \Gamma_{\Delta v90} ~ b)
+ \Gamma_{\rm [M/H]} ~ b + \beta - ze(z)
\end{equation}

\citet{wolfe1998} asserted that a \dv90-[M/H] relation existed, but
did not provide any quantification of its slope $\alpha_0$.
\citet{ledoux06} argued that a sample with a large internal scatter
like this was best fit with a bisector fit and reported 
$\alpha_0=1.55\pm0.12$. The non-homogeneous distribution of redshifts in
the their sample introduced a slight bias in the slope and after
correcting for $ze(z)$ \citet{moller13} found that $\alpha_0=1.46$
provided a good fit with the bisector method but that a minimum
$\chi^2$ fit would result in a flatter slope, $\alpha_0=1.12$.  
\citet{neeleman13} used a multi-parameter combined minimization
method and found $\alpha_0=0.74\pm0.21$ using a different
$ze(z)$ function. In summary, because of the large internal scatter
the reported slope of the main relation, $\alpha_0$, depends on the
fitting method used. However, as shown by \citet{moller13} the
determination of the redshift evolution of the relation is largely 
unaffected by the assumed slope of the main relation while the 
intercept, $\beta$, obviously is strongly correlated with the slope. 
Here we adopt the value $\alpha_0=1.46$ and $ze(z)$ as found by
\citet{moller13} which provides internal 
consistency with the results by \citet{christensen14} and 
\citet{arabsalmani15} to be used below. \citet{christensen14} found
$\Gamma_{\rm [M/H]}=-0.022$ dex \pkpc, and we can now write
\begin{equation}
\label{eq6}
{\rm [M/H] = 1.46} (\log(\dv90) - \Gamma_{\Delta v90} ~ b)
- 0.022 ~ b + \beta - ze(z)
\end{equation}
We note in passing that we also, as consistency test, computed the
metallicity gradient of the sample listed in Table~\ref{tab:sample}
and found $\Gamma_{\rm [M/H]}=-0.030\pm0.008$ dex \pkpc, consistent with the
values reported by \citet{christensen14} and \citet{rhodin2018}.

\begin{figure}
\includegraphics[width=8.5 cm]{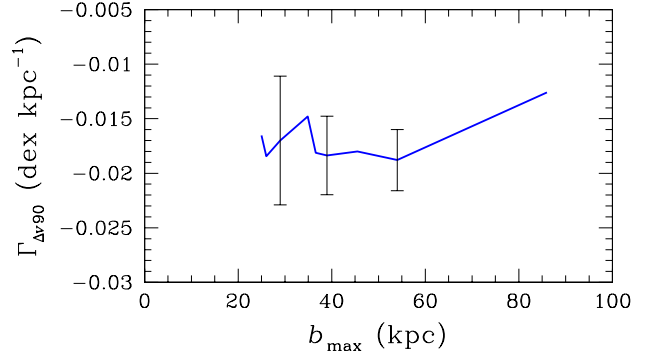}
\vskip -3.6 cm
\caption{The gradient $\Gamma_{\Delta v90}$ of $\log(\dv90)$ measured
from the centre of the DLA galaxy to $b_{\rm max}$ as a function of 
$b_{\rm max}$. It is seen that the gradient remains at a constant
value of $\approx -0.017$ dex \pkpc out to $b\approx60$ kpc.
$1\sigma$ errors on $\Gamma_{\Delta v90}$ for three 
representative values of $b_{\rm max}$ are also shown.
}
\label{fig:slopeFig}
\vskip -0.2 cm
\end{figure} 

\subsection{Quantifying $\Gamma_{\Delta v90}$}
\label{sect:Quantifying}

There are two ways that we can attempt to determine 
$\Gamma_{\Delta v90}$.
The first is to use the result from \citet{arabsalmani15} which says
that there should be no dependence on $b$ in equation~\ref{eq6}. In 
order to force the terms with $b$ to cancel, $1.46~\Gamma_{\Delta v90}$ 
must be equal to $-0.022$, i.e. $\Gamma_{\Delta v90}=-0.015$ as already 
shown in Sect.~\ref{sect:slopetest}. This result is based on three 
independent data samples: (1) the
GRB host sample providing [M/H], \dv90\ and \zabs\ to determine the
redshift corrected relation, (2) the larger DLA sample providing [M/H],
\dv90\ and \zabs\ for the comparison, and (3) the smaller DLA host 
sample providing $b$, ${\rm M}_*$, [M/H], \dv90\ and \zabs\ used to 
determine $\Gamma_{\rm [M/H]}$.

The other method is to measure $\Gamma_{\Delta v90}$ directly on the new
data provided here,
i.e. to fit a line to the data points of log(\dv90 / \sigmaem) vs. $b$ 
in Fig.~\ref{fig:MWtest} and determine its slope.
The DLA galaxies in our sample have, by necessity,
a large overlap with those of the \citet{christensen14} sample, but the
use of direct measurements of \sigmaem\ is new here. More importantly,
while both the MZ and the \dv90-[M/H] relations evolve with redshift,
and therefore indirectly are linked to the choice of $\alpha_0$ and
$ze(z)$, we here assume that \sigmaem\ and \dv90\ are purely dictated 
by gravity and as such have no specific dependence on $z$ and therefore 
also no links to choices already made.

In Sect.~\ref{sect:slopetest} we already presented fits to the QSO-DLA
data alone. Here we include the GRB-DLA data. We again discard
GRB~090323A, but including it will not change the conclusions. We are 
then left with a total sample of 21 QSO-DLAs and 9 GRB-DLAs for which 
we again assign the median GRB impact parameter of 1.0 kpc to each.
We use the method described in \citet{moller13} where we determine
both the optimal $\chi^2$ fit and the natural (intrinsic) scatter 
($\sigma_{\rm nat}$) simultaneously, and in order to test for a
flattening of the slope we perform the fit from $b=0$
kpc out to $b=b_{\rm max}$ for values of $b_{\rm max}$ from 25 to 86 
kpc. The result is shown in Fig.~\ref{fig:slopeFig} and we see that out 
to $b_{\rm max}=54$ kpc there is no evidence for any flattening as
$\Gamma_{\Delta v90}$ consistently displays a value of $\approx -0.017$
dex \pkpc. Only the object with the largest impact parameter at 86 kpc 
deviates from this trend, in agreement with the visual impression from 
Fig.~\ref{fig:MWtest}. The $1\sigma$ error on the fit to 29 objects 
within $b_{\rm max}=54$ kpc is 0.0028 dex \pkpc\
confirming the detection of the slope at $6 \sigma$. The corresponding
scatter is $\sigma_{\rm nat}=0.22$. The values of $\Gamma_{\Delta v90}$
from the two methods agree to within less than $1 \sigma$.

\subsection{Flattening of the gradients}

From the models of DM halos, as well as the observations of extended
flat rotation curves, it is expected that the steep CGM gradient we have
found for $\log(\dv90)$ should flatten a larger radii. We show this in
Fig.~\ref{fig:MWtest} as an example DM profile, and it is also
there seen that the object with the largest $b$ does not follow the
steep slope. It is slightly surprising that all the other points
agree well with the same steep slope as far out as 54 kpc (as also shown
in Fig.~\ref{fig:slopeFig}), but the sample is small with a significant
scatter, so it is premature to speculate about implications of this yet.

The fact that GRB hosts follow QSO-DLAs is interesting though. This
result is more significant since it is a single offset based on
larger samples. What this means is that if the $\log(\dv90)$ gradient
flattens at larger radii, then the metallicity would likely have to
behave in a similar way to remain consistent. In this sense, the
halo profile flattening would result in a prediction that there
must be a corresponding flattening of the metallicity gradient.

\subsection{Interpreting the \dv90-[M/H] relation}
\label{sect:Interpreting}

Following the previous sections, and recalling the errors on the two
gradients: $\Gamma_{\rm [M/H]}=-0.022\pm0.001$ and
$\Gamma_{\Delta v90}=-0.017\pm0.003$ we can now formulate the final
description of the \dv90-[M/H] relation (equation~\ref{eq6}) as
\begin{equation}
\label{eq7}
{\rm [M/H]} = 1.46 \log(\dv90) + (0.003\pm0.005)~b + \beta - ze(z)
\end{equation}
where we see that after determining the two gradients independently,
the dependence on impact parameter disappears to within the error.
This means that the hope that some of the scatter in the relation was
caused by a dependence on $b$, which could then be identified and
removed \citep{moller13}, has been foiled. The scatter must have other 
causes. This conclusion is based on the assumption that the two 
gradients remain
constant, or that they both flatten in a similar way. If only one of
them flattens, or is truncated, there could still be a $b$ dependence
in equation~\ref{eq6}, even though this seems to be ruled out by the
comparison to GRB hosts. An expansion of the current DLA galaxy samples
at larger impact parameters would be required to answer this.

The main two takeaway points concerning the \dv90-[M/H] relation is
therefore that the main slope, $\alpha_0$, still represents the
relation which exists for the centres of the galaxies, but that all
the individual galaxies have been pushed down along this relation to
a lower position, meaning that the metallicity follows the {\it local}
value of \dv90. The exact distance each galaxy has been pushed is
a function of $b$. For DLA galaxies we find a median in our sample of 
$b_{\rm med}=11.8$ kpc, which means that on average a DLA galaxy in 
the mass range considered here will be lying on the underlying
relation, but at a metallicity which is 0.26 dex lower than its
central value. Because sub-DLAs usually have larger impact parameters
they could have been shifted even further down the relation but if
the gradients flatten at some $b$ then that would define the maximum
shift.

The prediction above that the metallicity gradient does not add
scatter to the \dv90-[M/H] relation can be tested directly.  
In Appendix B we assemble an additional DLA galaxy sample for this
test.

\subsection{Quantifying $\beta$ and $ze(z)$}

Equation~\ref{eq5} provides the general form of the \dv90-[M/H] relation
as a relation through the centre of galaxies but modified by their
actual impact parameters. An observed relation reflects the same central
relation but then modified by the sample average of random impact 
parameters which in general would cause a modified $\beta$ and possibly
also a modified $ze(z)$. However, in Equation~\ref{eq7} we have showed
that the combined effect of the impact parameter is to move galaxies 
along the central relation which means that the observed relation is 
identical to the central relation, but all galaxies are instead pushed 
to lower
metallicities. $\beta$ and $ze(z)$ are therefore conserved within the
redshift range considered here, and for consistency with $\alpha_0$ and
$\Gamma_{\rm [M/H]}$ one must use the corresponding values
 $\beta = -3.33$ and $ze(z)=0.35~z$ (for $z<2.62$);
$ze(z)=0.35\times2.62$ (for $z>2.62$)
\citep{moller13}.

\subsection{A simple physical interpretation of the MZ relation}

Let us try to view the ``local history of the universe'' from the point
of view of local gravitational well depth. At a given gravitational
potential contour, and summed over the history of the universe, a 
certain amount of pristine gas will have passed this contour on its 
way into the well, a certain amount of metals will have been produced 
by stars inside the contour and a certain amount of 
enriched gas will have 
been pushed back out through this contour by outflows. Together those
three processes define the metallicity {\it on} the contour. 

The new results presented in this paper show that the way those
three processes (infall, star formation and outflow) set the local
metallicity, is directly linked to the local potential irrespective 
of redshift (in the redshift range covered by our data). In other 
words, the classic MZ relation is a special 
projection which is relating the integrated stellar mass inside the
potential well to the luminosity-weighted average metallicity of the galaxy.
The underlying relation is
directly between the local values of the potential
and the metallicity.

\section{Conclusions}

\citet{arabsalmani15} put forward the following bold proposal concerning
the root of the mass metallicity relation:
{\it `` ... then this means that the general
concept of an MZ relation plus metallicity gradients simply is a
convoluted and roundabout way of describing a much simpler underlying
relation between metallicity and gravitational well depth."}

In this paper we test this hypothesis. We carefully search the
literature and data archives, and find that the required observational
data for the test currently is at hand for 21 DLA/sub-DLA galaxies and
9 GRB host galaxies. The test predicts that out to at least the maximum
range of impact parameters for DLA galaxies, typically taken to be 20-30
kpc, the velocity dispersion of the halo gas should decrease as an
approximately linear relation with a slope of $-0.015$ dex kpc$^{-1}$.
We show that the current sample fully follows this prediction out to
impact parameters of 40-60 kpc. We further explore if there is any
evidence that this slope is a function of redshift, stellar mass or
\ion{H}{i} column density.
We also test if the ``classic dynamical tracer'', the relative velocity
between \zabs\ and \zem\ (in this paper named \vrel), holds similar
dynamical information as the absorption line velocity width, \dv90.
We find that this is not the case and argue that this can be used
to constrain the nature of individual DLA absorbing clouds.

Concisely our main findings can be summarized as
\begin{itemize}
\item
We confirm that the mass metallicity relation of DLA and GRB host
galaxies is local in nature. In particular that the metallicity (at
least out to 20-30 kpc but possibly further) directly traces the
gravitational potential.
\item
We show that the ``CGM halo potential'' of DLA galaxies, which here
is taken to mean the inner 40-60 kpc, to within the intrinsic scatter
can be described well by a linear function with a slope of
$\Gamma_{\Delta v90}=-0.017\pm0.003$ dex kpc$^{-1}$.  
\item
Our sample covers redshifts from $z=0.16$ to $z=3.15$, we see no change
in the above results over this redshift range.
\item
Our sample covers galaxy stellar masses in the range
$\log {\rm M}_*/{\rm M_\odot}=9.4-11.06$, we see no significant
evidence for a change in the above results over this stellar
mass range.
\item
We compare the dynamical tracers \dv90\ and \vrel, and find that taken
together they favour an interpretation that DLA absorption systems
are composed of a series of independent systems distributed along the
pencil beam through the halo, and disfavours a model of a single but
complex cloud moving with a bulk velocity.
\end{itemize} 

Based on the new results presented here it appears that the
\dv90-metallicity relation forms the underlying physical relation of 
which the well known MZ relation is a special projection, relating only 
the stellar mass fraction of the potential to the luminosity-weighted 
average metallicity of the galaxy.

\section{Acknowledgements}
It is a pleasure to thank Hadi Rahmani for kindly sharing HIRES and
MUSE spectra. We are grateful to our referee who made several
constructive suggestions. LC is supported by YDUN grant DFF 4090-00079. 
This is based on archival observations at ESO; Programme IDs:
074.A-0597(UVES),
078.A-0003(UVES),
078.A-0646(ADP),
278.A-5062(ADP),
080.A-0742(SINFONI),
087.A-0022(X-shooter)
and 
087.A-0414(X-shooter).

\def\aj{AJ}
\def\araa{ARA\&A} \def\apj{ApJ} \def\apjl{ApJ} \def\apjs{ApJS}
\def\apss{Ap\&SS} \def\aap{A\&A} \def\aapr{A\&A~Rev.}
\def\aaps{A\&AS} \def\mnras{MNRAS} \def\nat{Nature} \def\pasp{PASP}
\def\aplett{Astrophys.~Lett.}

\bibliographystyle{mn}
\bibliography{LiseSigma.bib}

\appendix 
\section{Extraction of sample data}
\label{app:a}

Below we provide details on individual absorbers and galaxies, including
references to relevant data. In cases where we measured \dv90\ or \ztau50\
on archival data (Q0021+0043, Q0153+0009, Q1439+1117, Q2233+1318, 
Q2328+0022 and Q2352-0028), we selected
appropriate metal line transitions that are not saturated following the
description in \citet{ledoux06}. Where we correct \dv90\ values for
resolution we use the method described by \citet{arabsalmani15}.

{\bf Q0021+0043 = SDSS J002133.27+004300.9}: \citet{peroux2016} report
$z_{\rm em}=0.94187$ from detection of \halpha\ emission at an impact
parameter of $b=10.8$ arcsec, corresponding to 86 kpc in our assumed
cosmology. They further report an emission line velocity dispersion of
$\sigmaem=123\pm11$ \kms.  \citet{rao2011} report $z_{\rm abs}=0.9420$
and $\log N($H\,{\sc i}$)=19.38_{-0.15}^{+0.10}$.  We measured \dv90\
= 139 \kms and $\ztau50=0.94181$ on the \ion{Si}{ii}(1808) line using
UVES ESO archive data (Programme-ID: 078.A-0003) \citep{peroux2016}.
We adopt \ztau50\ as \zabs\ and find $\vrel=-9\pm5$ \kms.

\begin{figure}
\hskip -0.6 cm
\includegraphics[width=8.8 cm]{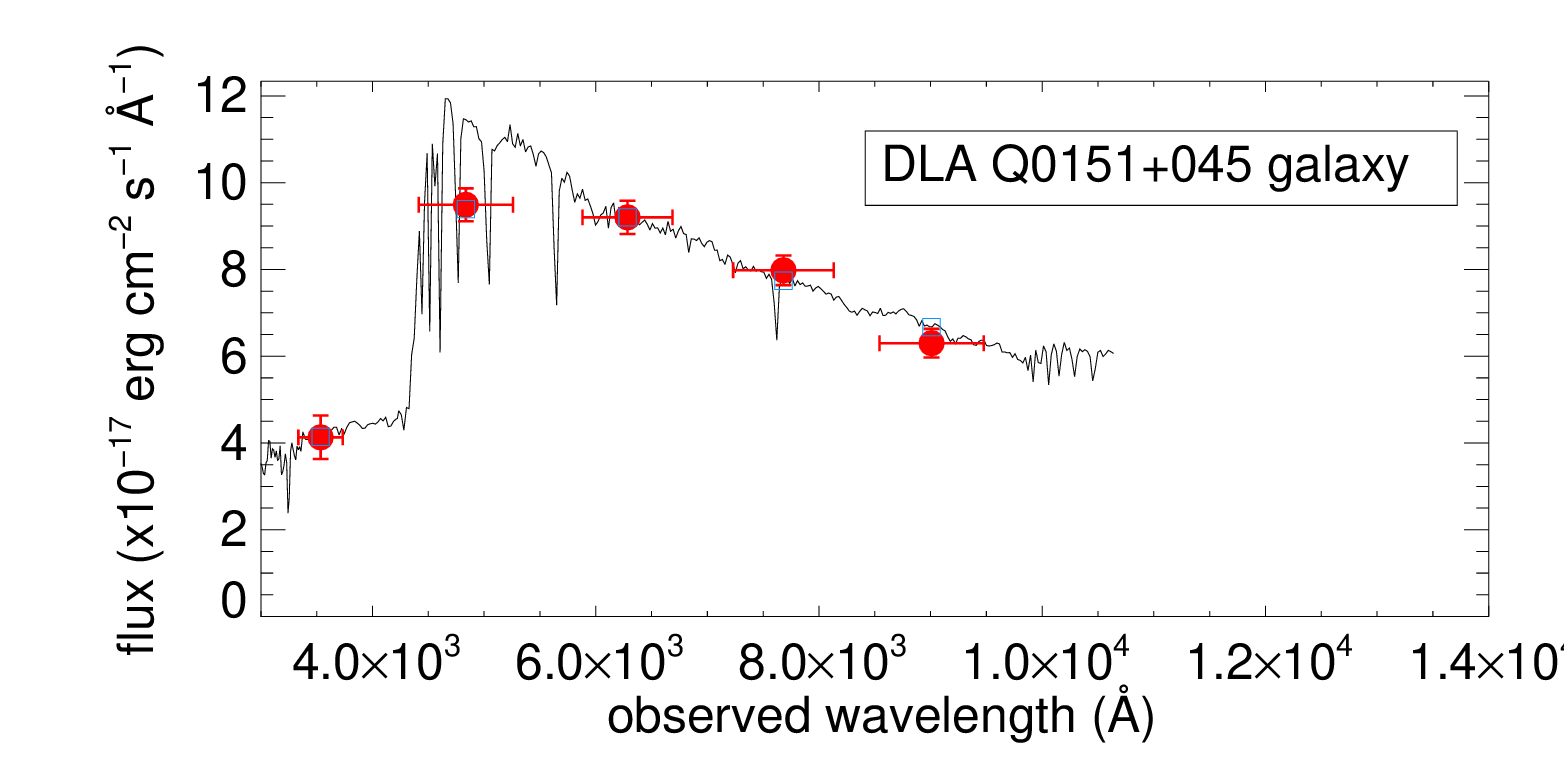}
\caption{
Spectral energy distribution fit from the HyperZ code
\citep{Bolzonella00} for PHL~1226 resulting in a stellar mass of log
M*$=9.73\pm0.04$ derived from fitting the data points to galaxy templates
using a Chabrier initial mass function.}
\label{fig:SEDPHL1226}
\end{figure}

{\bf Q0151+045 = PHL 1226 = J015427.99+044818.2}:
\citet{christensen2005} used IFU (PMAS) to identify the host of this
sub-DLA. The host (named G4) as seen in continuum is 6.5" south and
1.5" west of the QSO, i.e. the angular distance of the galaxy
continuum image is 6.7" and $b=18.5$ kpc. They also detected \halpha\
emission from G4 but with a centroid offset 1.4" to the south of the
continuum image (their Fig. 5). They report $\zem=0.1595\pm0.0006$,
but the line-emission appears to be kinematically coupled to only one
side of G4 (their Fig.7) and they find it is offset to a lower
redshift than the absorption by $\vrel=+180$ \kms . We have determined
\sigmaem$=50\pm20$ \kms\ (corrected for resolution) from the \halpha\
on the original data. \citet{som2015} reported $\log N($H\,{\sc
i})$=19.48\pm0.10$, $z_{\rm abs}=0.1602$ and \dv90$=152$ \kms. Using
SDSS photometry we have used standard SED fitting (see
Fig.~\ref{fig:SEDPHL1226}) and determined $\log {\rm M}_*/{\rm
M_\odot}=9.73\pm0.04$.

{\bf Q0152-2001 = J015227.32--200107.0}: using MUSE \citet{rahmani2018}
identified the host at $\zem=0.78025\pm0.00007$ of the
$\log N($H\,{\sc i}$)=19.1\pm0.3$ absorber.  They detect
[\ion{O}{ii}], [\ion{O}{iii}] and \hbeta\ emission lines in the host
at $b=54$ kpc, and from their HIRES spectrum they report $\vrel=-78$
\kms . They did not extract \sigmaem\ and \dv90, but they kindly
provided us with a copy of both the reduced MUSE and HIRES data. We
find $\sigmaem=104\pm13$ \kms (after correction for the spectral
resolution of 2.6 \AA ) and $\dv90=33$ \kms .

{\bf 0153+0009 = SDSS J015318.19+000911.4}: \citet{rhodin2018} report
$\zem=0.77085\pm0.00003$, $\sigmaem=121\pm8$ \kms
, $\log {\rm M}_*/{\rm M_\odot}=10.03^{+0.18}_{-0.08}$, and impact
parameter $b=36.6$ kpc of a candidate counterpart of the $z=0.7714$,
$\log N($H\,{\sc i}$)=19.70\pm0.09$ absorber \citep{rao2011}.  We
measure $\dv90=58$ \kms\ and $\ztau50=0.77219$ from UVES archive data
(Adv. Data Products, Programme ID: 078.A-0646) using the
\ion{Al}{iii}1854 line \citep{peroux2008}.  We adopt \ztau50\ as
\zabs\ and find $\vrel=+227\pm5$ \kms .

{\bf 0302-223 = J030449.87--221151.9}: \citet{peroux2011b} determine
$\zem=1.00946$, $\sigmaem=59$ \kms , $\vrel=-1$
\kms\ and impact parameter $b=25$ kpc of the counterpart of a $\log
N($H\,{\sc i}$)=20.36\pm0.11$ DLA at $\zabs=1.00945$
\citep{pettini2000}. They do not provide an error on \sigmaem
. Looking at their high S/N data we estimate that 10\% is a
conservative upper limit, and we therefore use $\sigmaem=59\pm6$.
\citet{moller13} determine $\dv90=61$ \kms\ and
\citet{christensen14} $\log {\rm M}_*/{\rm M_\odot}=9.65\pm0.08$.

{\bf 0918+1636-1 = SDSS J091826.16+163609.0}: for the $\log N($H\,{\sc
i}$)=21.26\pm0.06$ DLA \citet{fynbo13} report the detection of
\fion{O}{iii} with FWHM = $50\pm12$ \kms (corrected for resolution and
corresponding to $\sigmaem=21\pm5$), at $\zem=2.4128\pm0.0002$ and at
an impact parameter of $b<2$ kpc. They also report $\vrel=-38\pm25$
\kms and $\dv90=350\pm2$ \kms . Applying the correction for the
X-shooter 45 \kms\ resolution provides an intrinsic $\dv90=344$ \kms .

{\bf 0918+1636-2}: for the DLA at $\zabs=2.5832$ \citet{fynbo13} report
$b=16.2\pm0.2$ kpc, $\dv90=295\pm2$ \kms\ and
$\log N($H\,{\sc i}$)=20.96$. As above we apply the correction
for resolution and get $\dv90=288$ \kms. Based on their detection of
\fion{O}{iii} in emission they find $\zem=2.58277\pm0.00010$,
$\vrel=+36\pm20$ \kms\ and $\mathrm{FWHM=252\pm23}$ \kms\ (corrected for
resolution), i.e. $\sigmaem=107\pm10$ \kms.
$\log {\rm M}_*/{\rm M_\odot}=10.33\pm0.08$ was reported by
\citet{christensen14}.  

{\bf 1009-0026 = SDSS J100930.47--002619.1}: \citet{peroux2011b}
determine the emission redshift $\zem=0.8864$, and impact parameter
$b=39$ kpc of the counterpart of a $\log N($H\,{\sc
i}$)=19.48^{+0.05}_{-0.06}$ absorber at
$\zabs=0.8866$, i.e. $\vrel=+32$ \kms . Based on their \halpha\
SINFONI data cube they describe a distribution of \sigmaem\ peaking at
190 \kms\ in the centre but only 60-70 \kms\ in the outer parts. Since
they do not provide an integrated value we downloaded the data from
the ESO archive (programme-ID: 080.A-0742) and integrated the total
\halpha\ line to obtain $\sigmaem=174\pm5$ \kms . \citet{meiring2009b}
determined $\dv90=94$ \kms , and \citet{christensen14} report $\log
{\rm M}_*/{\rm M_\odot}=11.06\pm0.03$.

{\bf J1135-0010 = SDSS J113520.39--001053.6}:
\citet{noterdaeme2012} detect \fion{O}{iii} and \halpha\ emission at an
impact parameter of $b=0.8$ kpc (when converted to our cosmology) from this
$\log N($H\,{\sc i}$)=22.10\pm0.05$, $\zabs=2.2066$, DLA, but they
provide only an approximate FWHM. We therefore use the ADS Dexter data
extraction applet \citep{demleitner2001} to measure individual values
of the three detected lines and combine them to obtain
$\sigmaem=53\pm3$ \kms\ (corrected for resolution).
They do not provide separate \zabs\ and \zem , but from their figure 3
we see that the individual transitions show a range of \vrel\ covering
both positive and negative values. On average we find $\vrel=-20\pm10$ \kms.
\citet{kulkarni2012} published high resolution spectroscopy but did
not measure \dv90 . We used ADS Dexter to extract $\dv90=168$ \kms\
using the unsaturated \ion{Cr}{ii}2056 line.

{\bf 1228-1139/B1228-113 = J123055.56--113909.8}: \citet{neeleman2018}
report the detection of line emission from CO and \halpha\ at an
impact parameter of 3.5 arcsec, corresponding to $b=30$ kpc at
$\zabs=2.19289$. They further find $\log N($H\,{\sc
i}$)=20.60\pm0.10$, $\dv90=163\pm10$ \kms , $\zem=2.1912$ (\halpha
). From this we find $\vrel=+159\pm18$ \kms , and from the extracted
SINFONI spectrum (programme-ID: 080.A-0742) we measure
$\sigmaem=93\pm31$ from \halpha .

{\bf B1323-0021 = J132323.78--002155.3}: \citet{moller2018} report the
detection of line emission from CO, \halpha\ and \fion{O}{ii} from a
candidate DLA galaxy \citep{hewett2007} at an impact parameter of 1.25
arcsec \citep{chun2010}. They further find $\log {\rm M}_*/{\rm
M_\odot}=10.80^{+0.07}_{-0.14}$, $b=9.1$ kpc, $\log N($H\,{\sc
i}$)=20.4^{+0.3}_{-0.4}$, $\zem=0.7171\pm0.0001$,
$\zabs=\ztau50=0.71612$ \kms , $\vrel=-171\pm18$, $\dv90=141\pm2$
\kms\ and $\sigmaem=101\pm14$ \kms (based on their FWHM of \halpha ).

{\bf 1436-0051A = SDSS J143645.05--005150.6}: \citet{rhodin2018}
determined $\zem=0.73749\pm0.00003$, $\sigmaem=99\pm25$ \kms ,
$\log {\rm M}_*/{\rm M_\odot}=10.41^{+0.09}_{-0.08}$, and $b=45.5$ kpc to
confirm a candidate counterpart of the $\zabs=0.7377$, $\log N($H\,{\sc
i}$)=20.08\pm0.11$, $\dv90=71$ \kms\ absorber \citep{meiring2009b}.

{\bf 1436-0051B} \citet{rhodin2018}: determined the emission redshift
$\zem=0.92886\pm0.00002$, $\sigmaem=33\pm11$ \kms , 
$\log {\rm M}_*/{\rm M_\odot}=10.20^{+0.11}_{-0.08}$,
and impact parameter $b=34.9$ kpc of a candidate counterpart of the
$\zabs=0.9281$, $\log N($H\,{\sc i}$)<18.8$, $\dv90=62$ \kms\ absorber
\citep{meiring2009b}. 
\citet{meiring2008} reported $\log N($Zn\,{\sc ii}$)=12.29\pm0.06$ and
\citet{straka2016} $\log N($H\,{\sc i}$)=18.4\pm0.98$. Without 
ionization correction this would provide a metallicity of 
$+1.26\pm1.02$. \citet{straka2016} computed ionization corrections and 
found a metallicity of $-0.05\pm0.55$ implying that the computed 
ionization corrections are the range $-1.78$ to $-0.84$ dex or a factor
of 60 to 7.

{\bf 1439+1117 = SDSS J143912.04+111740.6}: \citet{rudie2017} report the
detection of line emission at $\zem=2.4189\pm0.0001$ from a $\log {\rm
M}_*/{\rm M_\odot}=10.74^{+0.18}_{-0.16}$ galaxy at impact 4.7 arcsec
($b=39$ kpc) from J1439+1117.  They give \vrel\ as $-46$ \kms\ and
$\sigmaem=303\pm12$ \kms.  We downloaded the UVES archive data used by
\citet{srianand2008} (Adv. Data Products, programme ID: 278.A-5062)
and measure $\dv90=338$ \kms\ and $\ztau50=2.41802$, both using the
\ion{Fe}{ii}1608 line. Using now \ztau50\ for \zabs\ we find
$\vrel=-77$ \kms. \citet{srianand2008} report $\log N($H\,{\sc
i}$)=20.10\pm0.10$. \citet{rudie2017} argue that the host of this
absorber is AGN dominated, and that the absorber therefore likely is
influenced by AGN outflow. We include it in our sample because this
provides us with an opportunity to test if a strong outflow will cause
absorption systems to diverge from standard kinematic behaviour or
not.

{\bf LBQS 2206-1958 = J220852.07--194359.86}: \citet{weatherley2005}
report \fion{O}{iii} line emission from two interacting galaxies at
$\zem=1.9220(2)$ and $\zem=1.91972(8)$, both associated to a DLA at
$\zabs=1.919991(2)$. The two galaxies have impact parameters of 0.98
and 1.24 arcsec respectively, corresponding to $b=8.4\pm0.3$ and
$b=10.6\pm0.3$ kpc, and their line widths and velocity offsets are
given as FWHM$=220\pm50$ \kms , FWHM$=180\pm25$ \kms ,
$\vrel=-210\pm20$ \kms\ and $\vrel=+29\pm9$ \kms .  \citet{ledoux06}
report $\dv90=136$ \kms\ and $\log N($H\,{\sc i}$)=20.67\pm0.05$. The
two galaxies are members of what appears to be an active merger
\citep{moller02} and \citet{christensen14} used SED fitting to compute
the total stellar mass of the entire group $\log {\rm M}_*/{\rm
M_\odot}=9.45\pm0.30$.  Here we assign ``ownership'' of the merging
group to the main galaxy (named N-14-1C) at $b=8.4\pm0.3$ kpc and
compute $\sigmaem=93\pm21$ \kms.

{\bf 2222-0946 = SDSS J222256.11--094636.3}: \citet{krogager13} report
the FWHM (corrected for resolution) of 5 different emission lines at
$\zem=2.3537$ and at good S-to-N. We combine all 5 measurements using
inverse variance weighting and find $\sigmaem=49.0\pm1.6$ \kms
. \citet{fynbo2010} report $\log N($H\,{\sc i}$)=20.65\pm0.05$ and
$\dv90=185$ \kms\ which after correction for resolution (45 \kms )
becomes 174 \kms. \citet{krogager17} find $b=6.3\pm0.3$ kpc and
\citet{christensen14} find $\log {\rm M}_*/{\rm M_\odot}=9.62\pm0.12$.
\zem\ is embedded inside the rather wide span of absorption
components.  To obtain a velocity offset we estimate \ztau50\ as follows.  
From Table 2
of \citet{krogager13} we find that for three of the six low ionization
species listed the median is in the second component (at 11 \kms )
while for the other three it is in the third component (at 73 \kms
). We therefore assign the relative offset to be between the two at
$\vrel=+42\pm30$ \kms.

{\bf 2233+1318 = J223619.20+132620.3}: \citet{weatherley2005} report
\fion{O}{iii} line emission at $\zem=3.15137\pm0.00006$ with
$\sigmaem=23^{+8}_{-13}$ \kms . They also compute the velocity offset
between \zem\ and $\zabs=3.14930(7)$ to be $-150$ \kms .  In their
section 3.1 they point out that the location of the line emission
precisely matches the position of an existing HST-STIS image, but not
the position of an HST-NICMOS image which is offset by 2.1
kpc. Therefore, this is likely line emission from a single site of
star formation in an otherwise quiescent galaxy and as such the emission
line width is representative of the smaller UV dominated region rather
than the general, and older, stellar population.  From the STIS
position we obtain $b=19.5$ kpc, using the NICMOS position we find
$b=21.6$ kpc.  \citet{christensen14} found $\log {\rm M}_*/{\rm
M_\odot}=9.85\pm0.14$.  We downloaded ESO-X-shooter archival data (PI:
Cooke, Programme-ID: 087.A-0022) and measured $\dv90=236.9$ \kms\
using the \ion{Fe}{ii}1608 line.  Correction for X-shooter resolution
provides an intrinsic $\dv90=228$ \kms.
We also measured the HI column density
($\log N($H\,{\sc i}$)=20.00\pm0.10$) and performed
multi-component Voigt profile fitting of \ion{Fe}{ii}1608 and
\ion{Si}{ii}1260,1304,1526,1808 using the code VoigtFit described in
\citet{krogager18}. Among the fitted lines only \ion{Si}{ii}1260 
appears to be saturated. We determined column densities
$\log N($Fe\,{\sc ii}$)=14.10\pm0.15$ and 
$\log N($Si\,{\sc ii}$)=14.54\pm0.08$ resulting in metallicities of 
[FeII/H]$ = -1.37\pm0.18$ and [SiII/H]$ = -0.97\pm0.13$.

{\bf 2243-6031 = J224709.10--601545.0}: \citet{bouche2013} detect
several emission lines at $\zem=2.3283\pm0.0001$ and report
$\sigmaem=158\pm5$ for a DLA host at an impact parameter of 3.1 arcsec
($b=26$ kpc). Based on a deep VLT/UVES spectrum \citet{lopez2002} had
previously reported $\log N($H\,{\sc i}$)=20.67\pm0.02$ for this
absorber and a low-ion spectral complex consisting of 14 individual
components spanning 330 \kms . From their fit to each component, and
using the same method as described for 2222-0946 above, we find that
\zabs , again defined here as the median optical depth redshift
\ztau50, is $2.3298\pm0.0001$ leading to $\vrel=+135\pm12$
\kms .  \citet{ledoux06} find $\dv90=173$ \kms\ while Rhodin (in
preparation) report $\log {\rm M}_*/{\rm M_\odot}=10.1\pm0.1$.

{\bf 2328+0022 = SDSS J232820.37+002238.1}: \citet{rhodin2018}
determined the emission redshift $\zem=0.65194\pm0.00006$,
$\sigmaem=56\pm24$ \kms , $\log {\rm M}_*/{\rm M_\odot}=10.62\pm0.35$,
and impact parameter $b=11.9$ kpc of the candidate counterpart of a
$\zabs=0.6519$, $\log N($H\,{\sc i}$)=20.32\pm0.06$ DLA
\citep{rao2011}. We measure $\dv90=92$ \kms\ and $\ztau50=0.65179$
using the \ion{Mg}{i}2852 line obtained from UVES archive data
(programme ID: 074.A-0597). This results in $\vrel=-27\pm11$ \kms .

{\bf 2352-0028 = SDSS J235253.51−002850.4}: \citet{peroux2013} report
detection of \halpha\ from the host of the sub-DLA with
$\log N($H\,{\sc i}$)=19.81^{+0.14}_{-0.11}$ at $\zabs=1.0318$
\citep{rao2006}.  We downloaded the raw Sinfoni archive data 
(programme ID: 085.A-0708(A)) and measured a velocity dispersion of
$125\pm6$ km~s$^{-1}$ by fitting the integrated \halpha\ emission 
line of the DLA galaxy, and correcting for instrumental resolution.
\citet{peroux2013} also provide (in their Fig. 10) a
direct overlay of emission redshift on the low ion absorption profiles
from which we estimate that \ztau50\ is $40\pm5$ \kms\ higher
than \zem , i.e. $\vrel=+40\pm5$ \kms . \citet{meiring2009b} report
$\dv90=164$ \kms , and \citet{augustin2018} find $\log {\rm M}_*/{\rm
M_\odot}=9.4\pm0.3$ and an impact parameter of 1.50 arcsec
corresponding to $b=12.2$ kpc.  We measure $\ztau50=1.03197$ from
X-shooter archive data (programme ID 087.A-0414) using the
\ion{Fe}{ii}2374 line.

{\bf 2358+0149 = SDSS J235854.4+014955.5}: \citet{srianand2016}
detected \fion{O}{iii} emission at an impact parameter of $1.5\pm0.1$
arcsec with an intrinsic FWHM of 110 \kms\ corresponding to \sigmaem\
= 46.7 \kms . They provide no error on the emission line width, but
from the figures we estimate an upper limit on the error of 20\% and
assign conservatively an error of 9 \kms. They report \zabs\ and \zem\
of 2.97919 and 2.9784 respectively resulting in a velocity difference
of $+60$ \kms. In our cosmology the impact parameter corresponds to
$b=11.8\pm0.8$ kpc. They also provide \dv90\ values for an
\ion{Fe}{ii} line and \ion{Si}{ii}1808. The \ion{Fe}{ii} line is saturated so
cannot be used, for the \ion{Si}{ii}1808 line they give 141 \kms\
which (after correction for resolution) provides an intrinsic value of
135 km/s. They find $\log N($H\,{\sc i}$)=21.69\pm0.10$.

\section{Metallicity gradient and scatter of the \dv90-[M/H] relation}
\label{app:b}

\begin{figure}
\hskip -0.3 cm
\includegraphics[width=8.8 cm]{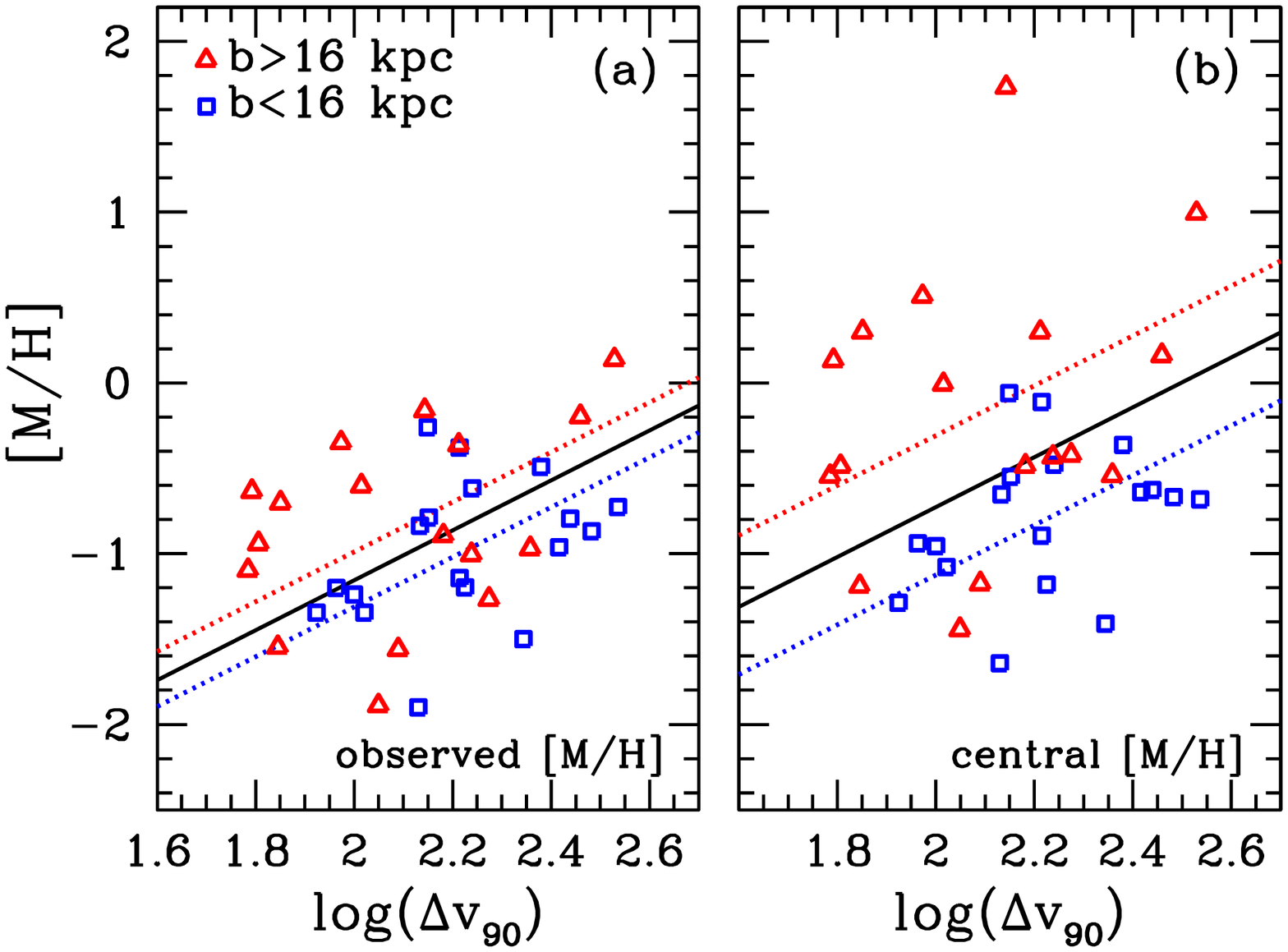}
\vskip -2.4 cm
\caption{Standard \dv90-[M/H] relation plots of the combined sample listed
in Tables~\ref{tab:sample} and \ref{tab:Xsample}.
The colour/symbol coding divides the sample into small and large impact
parameter. (a): observed metallicities only corrected for redshift
evolution. (b): metallicities computed for the centres of the galaxies
assuming a constant metallicity gradient independent of impact
parameter.  As predicted there is a larger scatter in (b).
}
\label{fig:FigPlusFig}
\end{figure} 

In Sect.~\ref{sect:slopetest} and again in Sect.~\ref{sect:Quantifying} 
we prove, using two different methods, that the \dv90 radial gradient exactly
cancels the effect of the metallicity gradient in the inner part of galaxy
halos. From this, and from equation~\ref{eq7}, we then assert in
Sect.~\ref{sect:Interpreting} that it must follow that the metallicity
gradient in the inner part of the CGM cannot significantly contribute to
the scatter of the \dv90-[M/H] relation. Here we test this assertion.
The test requires knowledge of the absorption metallicity of each galaxy,
which is only the case for 19 of the galaxies in our original sample.
However, the test does not require knowledge of \sigmaem\ which makes it
possible to expand our statistical sample for this test by including
DLA/sub-DLA galaxies for which \sigmaem\ is not known. The relevant data
for this additional sample are listed in Table~\ref{tab:Xsample} where 
we again have recomputed metallicities using \citet{asplund2009} and the
choices of \citet{decia2016} (their Table 1).

\begin{table*}
\caption{Additional literature sample of DLA/sub-DLA galaxies for which
impact parameter, \dv90 and absorption metallicity is known, but where
\sigmaem\ is not known. }
\begin{tabular}{lllccrrc}
\hline
ID & \zabs\ & $\log N($\ion{H}{i}) & [M/H]  & Tracer & $b$ & \dv90\ & References \\
   &        &                      &        &        & kpc & \kms\  &       \\
\hline
PKS~0439-433& 0.1012 &$19.63\pm0.15$ &$+0.08\pm0.15$ & S  &  7.6 & 275 & ~~1, ~~1, ~~1, ~~2, ~~1\\
1127-145    & 0.3127 &$21.71\pm0.08$ &$-0.76\pm0.10$ & Zn & 17.5 & 123  & ~~3, ~~3, ~~4, ~~5, ~~4 \\
0827+243    & 0.5247 &$20.30\pm0.04$ &$-0.54\pm0.05$ & Fe+0.4 & 38.4 & 188  & ~~4, ~~4, ~~4, ~~5, ~~4\\
0218-0831   & 0.5899 &$20.84\pm0.12$ &$-0.26\pm0.15$ & Zn & 14.7 & 261  & ~~6, ~~6, ~~6, ~~6, ~~6\\
1138+0139   & 0.6130 &$21.25\pm0.10$ &$-0.65\pm0.11$ & Zn & 12.2 & 105  & ~~6, ~~6, ~~6, ~~6, ~~6\\
0958+0549   & 0.6557 &$20.54\pm0.15$ &$-1.21\pm0.18$ & Zn & 20.3 & 112  & ~~6, ~~6, ~~6, ~~6, ~~6\\
2335+1501   & 0.67972&$19.70\pm0.30$ &$+0.07\pm0.34$ & Zn & 27.0 & 103.5& ~~7, ~~7, ~~7, ~~8, ~~9\\
0452-1640   & 1.0072 &$20.98\pm0.07$ &$-0.99\pm0.08$ & Zn & 16.2 &  70  & 10, 10, 10, 11, ~~9\\
Q1313+1441  & 1.7941 &$21.30\pm0.10$ &$-0.86\pm0.14$ & Zn & 11.2 & 164 & 12, 12, 12, 12, 12\\
2239-2949   & 1.82516&$19.84\pm0.14$ &$-0.67\pm0.15$ & Si & 20.6 &  64 & 13, 13, 13, 13, 13\\
PKS0458-02  & 2.0396 &$21.65\pm0.09$ &$-1.15\pm0.10$ & Zn &  2.7 &  84 & 14, 14, 14, 12, 14\\
0338-0005   & 2.229  &$21.12\pm0.05$ &$-1.37\pm0.06$ & Si &  4.12& 221 & 12, 12, 12, 12, 12\\
0124+0044   & 2.261  &$20.70\pm0.15$ &$-0.67\pm0.16$ & Zn & 10.9 & 142 & 15, 15, 15, 16, 15\\
Q2348-11    & 2.4263 &$20.53\pm0.06$ &$-0.43\pm0.08$ & Zn &  5.8 & 240 & 12, 12, 12, 12, 12\\
0139-0824   & 2.6773 &$20.70\pm0.15$ &$-1.24\pm0.20$ & Si & 13.0 & 100 & 16, 16, 17, 17, 16\\
0528-250    & 2.8110 &$21.35\pm0.07$ &$-0.87\pm0.07$ & Zn &  9.14& 304 & 18, 18, 18, 19, 18\\
\hline
\hline
\end{tabular}
\flushleft
References:
(1) \citep{som2015} the metallicity [S/H] includes their ionization 
correction of -0.18 dex;
(2) \citet{chen2005}; 
(3) \citet{lane1998};
(4) \citet{kanekar2014}, for 0827+243 they estimate [M/H] as [Fe/H]+0.4, for
details see footnote ``$c$'' to their Table 4;
(5) \citet{christensen14};
(6) \citet{rahmani2016};
(7) \citet{meiring2009b};
(8) \citet{rhodin2018};
(9) This work, measured on UVES archive data;
(10) \citet{peroux2008};
(11) \citet{augustin2018};
(12) \citet{krogager17};
(13) \citet{zafar2017};
(14) \citet{berg2015};
(15) \citet{berg2016};
(16) \citet{wolfe2008};
(17) Rhodin etal. in prep;
(18) \citet{ledoux06};
(19) \citet{moller02}.
\label{tab:Xsample}
\end{table*}

In Fig.~\ref{fig:FigPlusFig}(a) vi plot the standard \dv90-[M/H] relation
for this sample, 
corrected for redshift evolution by shifting all metallicities to $z=2.6$ as
described in \citet{moller13}. The sample contains 35 objects with a median
impact parameter of 16 kpc. Objects with $b>16$ kpc are shown as red
triangles, objects with smaller $b$ are marked as blue squares.
In Fig.~\ref{fig:FigPlusFig}(b) we plot the same
objects but here we show instead the central metallicities as computed from
the impact parameters and assuming a constant metallicity gradient
throughout. It is easy to see that this process in fact has increased the 
scatter rather than reduced it. To quantify this we have computed the usual
linear relations to both the red and blue sub-samples (dotted lines) as well
as to the total sample (full black line) in both figures B1(a) and B1(b). We
use the fitting method from \citet{moller13} which computes zero point and
intrisic scatter for a slope of 1.46 (see Sect.~\ref{sect:The_detailed}).
The zero point offset between the red and blue sub-samples grows from 0.32
dex to 0.82 dex and the intrinsic scatter of the fit to the full sample
grows from 0.46 dex to 0.72 dex when the central metallicities are used.
This confirms the prediction that ``correcting'' for only the effect of
metallicity gradient does not reduce the scatter of the relation as one
would expect if \dv90 did not depend on impact parameter.

\end{document}